\documentstyle[12pt,epsf]{article}		

\def\ion#1#2{#1$\;${\small\rm{#2}}\relax}

\def\beq{\begin{equation}}
\def\eeq{\end{equation}}
\def\beqar{\begin{eqnarray}}
\def\eeqar{\end{eqnarray}}

\def\pref#1{(\ref{#1})}   

\def\la{\mathrel{\mathpalette\fun <}}

\def\fun#1#2{\lower3.6pt\vbox{\baselineskip0pt\lineskip.9pt
  \ialign{$\mathsurround=0pt#1\hfil##\hfil$\crcr#2\crcr\sim\crcr}}}

\def\iso#1#2{\hbox{${}^{#2}{\rm #1}$}}
\def\he#1{\iso{He}{#1}}
\def\li#1{\iso{Li}{#1}}
\def\be#1{\iso{Be}{#1}}
\def\b1#1{\iso{B}{1#1}}
\def\feh{\hbox{$[{\rm Fe/H}]$}}

\def\lip{\hbox{$\li{7}_{\rm p}$}}

\def\dofe{\hbox{$\omega$}}
\def\nupro{$\nu$-process}

\begin{document}
\rightline{UMN--TH--1762/99}
\rightline{TPI--MINN--99/25}
\vskip 1in
\baselineskip 18pt

\begin{center}
{\Large\bf Primordial Lithium and Big Bang Nucleosynthesis}
\end{center}
\vskip 10mm


\vskip 10mm
S. G. Ryan$^*$,
T. C. Beers$^\dagger$,
K. A. Olive$^\ddagger$,
B. D. Fields$^{\S}$,
J. E. Norris$^{||}$

\vskip 05mm
* {\it Physics Department, Open University, Walton Hall, Milton Keynes MK7 6AA, United Kingdom}

$\dagger$ {\it Department of Physics and Astronomy, Michigan State University, East Lansing, MI 48824, USA}

$\ddagger$ {\it Theoretical Physics Institute, School of Physics and Astronomy, University of Minnesota, \\ 116 Church Street, SE, Minneapolis, MN 55455, USA}

\S {\it Department of Astronomy, University of Illinois at Champaign-Urbana,\\ 1002 West Green Street, Urbana, IL 61801, USA}

$||$ {\it Research School of Astronomy and Astrophysics, The Australian National University, Private Bag, Weston Creek Post Office, ACT 2611, Australia}

\vfill
\eject

{\bf
In the standard hot big bang nucleosynthesis (BBN) 
model$^{\cite{st98,fkot}}$, the primordial abundances of $^1$H, $^2$H, $^3$He, 
$^4$He, and $^7$Li fix the baryon density of the universe, $\Omega_b$, via the 
baryon-to-photon ratio, $\eta$, for a given Hubble parameter. 
Recent observations
of Li show$^{\cite{rnb99}}$ that its intrinsic dispersion in metal-poor stars is
essentially zero, and the random error in the mean Li abundance is negligible. 
However, a decreasing trend in the Li abundance towards lower metallicity,
plus
$^6$Li detections$^{\cite{sln98,cssvc99}}$, 
indicate that its primordial abundance can 
be inferred only after allowing for nucleosynthesis processes in the Galaxy
(Galactic chemical evolution, hereafter GCE). We show 
that the Li {\it vs} Fe trend provides a tough discriminant between alternative
models for GCE of light-elements. We critically assess current systematic 
uncertainties, and determine the primordial Li abundance within new, much 
tighter limits. We show that the Li constraint on $\Omega_b$ is now limited 
principally by uncertainties in the nuclear cross-sections used in BBN 
{\it calculations}, not by the abundance itself. 
A clearer understanding of systematics 
allows for a much more accurate inference of the primordial Li abundance, 
sharpening the comparison with \he4 and deuterium and
the resulting test of BBN.
We find that the Li data are in good agreement 
with $^4$He and with ``high'' deuterium values, but that low deuterium 
abundances are at best marginally within the Li range.

}

An important test of BBN is concordance between the observationally-inferred 
primordial abundances of the light elements. However, all current values 
involve considerable systematic uncertainties.
Estimates of the $^4$He primordial mass fraction, $Y_p$, had 
settled$^{\cite{p}}$ around $Y_p~=~0.230\pm~0.005$, but different systematics 
and underlying stellar \ion{He}{I} absorption may imply a higher 
value$^{\cite{it98}}$ near 0.245. For deuterium, quasar absorption line 
measurements give both ``low''$^{\cite{bt98,bt98a}}$ abundances around 
D/H~=~(3--5)$\times 10^{-5}$ and ``high''$^{\cite{schr94,crwcw94}}$ values 
around D/H~=~(15--25)$\times 10^{-5}$. It is unclear which value (if either)
represents the correct {\em primordial} value.
$^3$He presents even greater difficulties associated with its highly uncertain
yield in low-mass stars$^{\cite{gpfp95,orstv}}$, and currently provides 
unreliable constraints. Recent observations$^{\cite{rnb99}}$ of Li have greatly
improved, therefore we examine the random and systematic 
uncertainties associated with the primordial abundance and its
interpretation with respect to BBN.

In inferring the primordial $^7$Li abundance, $A$(Li)$_p$
(where $A$(Li) ~ $\equiv$ ~ log$_{10}(n$(Li)/$n$(H)) ~+~12.00),
from observations of metal-poor stars, systematic errors arise
in: (1) the assessment of Li GCE prior to a given star forming;
(2) the correction for depletion of a star's initial surface Li; 
(3) the measurement of the current abundances; and
(4) possible confusion by anomalous objects.
We examine each factor below, and summarise its impact in Table~1.

{GCE:} 
The GCE contribution to $^7$Li was long believed to be negligible for
metal-poor stars. However, observations of very metal-poor stars showing
(1) lower Li abundances$^{\cite{rnb99}}$ and   
(2) measurable $^6$Li$^{\cite{sln98,cssvc99}}$  
(which comes solely from Galactic
cosmic ray (GCR) reactions), show that GCE cannot be ignored. Li GCE can be 
constrained either empirically from the observed evolution, or by modeling the
sources and sinks of the element. Here we pursue both routes, partly to 
investigate the range of possible values, and partly to learn about GCE itself,
since an accurate Li trend can constrain not only standard GCR and 
\nupro\ nucleosynthesis, but also other mechanisms which may produce Li in 
Population II, e.g., those suggested to produce primary beryllium 
along with Li.
Sources of lithium in the oldest (Population II) stars are GCR nucleosynthesis 
of \li6 and \li7, and the supernova \nupro\ which produces \li7 and \b11. 
The sink is astration --- the destruction of
Li in stellar interiors subsequently ejected into the interstellar medium (ISM).

Previously$^{\cite{rnb99}}$ we examined GCE empirically, tracing the production
of Li as the iron abundance increased, as a regression in logarithmic 
abundances:
\beq
\label{eq:log_fit}
A({\rm Li}) = \alpha + \beta {\rm [Fe/H]}
\eeq
where 
[Fe/H]~$\equiv~\log_{10}({\rm Fe/H})_{\rm star} -
   \log_{10}({\rm Fe/H})_{\rm sun}$. 
(Iron is a convenient and widely-measured
diagnostic of GCE. It is always produced and never destroyed by stars, 
and thus serves as a chronometer.) We obtained values in the range 
$\beta~=~0.07$--0.16, depending on the adopted [Fe/H] values, the errors, and 
whether some points are excluded from the fit. 

In the present work, we also investigate 
a fitting form which better follows Li production by GCE,
and which simplifies the extrapolation to the primordial value.
Li production is proportional to the cumulative number, $N_{SN}$, of Type II 
supernovae, as these are both GCR accelerators and the site of the \nupro.
It thus
is important to establish the primary tracer of such supernovae. 
If the cumulative supernova rate is well reflected by the iron abundance
($N_{SN} \propto$ Fe), then a fit to {\it linear} 
abundance scales is appropriate:
\beq
\label{eq:Fe_fit}
{\rm Li/H} = a' + b' {\rm Fe/Fe_\odot}
\eeq
Here $a'$ directly measures the primordial \li7 abundance 
(in the absence of systematic errors), 
while $b'$ probes GCE. 
The linear fit parameters are sensitive to systematic Li 
errors; a change by $\Delta_{\rm cal}$~dex in the log shifts both $a'$ {\it and}
$b'$ by a factor $10^{\Delta_{\rm cal}}$. We find 
$a'~=~1.0$--1.2 $\times 10^{-10}$ and 
$b'~=~40$--180 $\times 10^{-10}$. 

On the other hand, if oxygen 
(which is more difficult to measure than iron) is a 
better tracer of Type II supernovae than iron, 
then $N_{SN} \propto$ O, and we expect:
\begin{equation}
\label{eq:lin_fit}
{\rm Li/H} = a + b \, {\rm O/O_\odot}
\end{equation}
where ${\rm O/O}_\odot = ({\rm Fe/Fe}_\odot)^{1+\omega}$. 
Recent observations$^{\cite{isr,boes}}$ show $\dofe = -0.31$. In this case the 
data indicate
$a~=~0.9$--1.2 $\times 10^{-10}$ and 
$b~=~9$--34 $\times 10^{-10}$.

We also compute the Li-Fe trend expected from a one-zone (closed box) GCE 
model$^{\cite{fo98,fo99,fo99na}}$ 
which includes GCR and stellar nucleosynthesis, to
compare with the observational results.  (See Figure 1 caption for details.)
Figure 1 
shows the different Li components for the model with 
$\lip = 1.23 \times 10^{-10}$. We fit the {\it model} by regressions of the forms
\pref{eq:log_fit} -- \pref{eq:lin_fit}, over the metallicity range of the
recent data$^{\cite{rnb99}}$, and find $b = 8.6 \times 10^{-10}$,
$b' = 4\times 10^{-9}$, and 
$\beta = 0.03-0.07$ (for input \lip = (0.9--1.9)$\times 10^{-10}$). 
(The model's ``linear slopes'', $b$ and $b'$, are independent of the input 
\lip, while its ``log slope'', $\beta$, does depend on \lip.)
This model's Li GCE slopes ($b$, $b'$ and $\beta$) are at the lower range of 
those found for the data, and for a closed-box model with only GCR and the 
\nupro. 
The poor agreement can be alleviated somewhat by manipulation of free 
parameters. 
For example, models including outflow of
supernova ejecta (open box) and/or the inclusion of Li yields from AGB stars 
with $M < 5 M_\odot$ produced steeper slopes, 
$b = (11-17)\times 10^{-10}$,
$b' = (6-9)\times 10^{-9}$, and
$\beta = 0.06-0.08$ (for various types of models with 
\lip = 1.23$\times 10^{-10}$), closer to the range seen in the data.
This demonstrates the power of a reliable observational Li {\it versus} Fe trend
to constrain the form and parameters of GCE models.

While comparison of model slopes with the observations can teach much
about GCE, to infer the primordial Li we use the observed
slopes (a procedure very similar to that used for \he4).
{}From the linear fits to the data and our previous analysis$^{\cite{rnb99}}$,
we estimate that the GCE contribution to this metal-poor turnoff sample is 
$-0.11^{+0.07}_{-0.09}$ in the log (see Table 1). 

{ Stellar Depletion:}
Stars burn Li, preserving at most a thin 
outer shell containing a few percent of the star's mass. Possible {\it in situ} 
depletion of Li has long been regarded the major systematic uncertainty in
inferring $A$(Li)$_p$ from present-day abundances. Stellar 
evolution models predict the depletion factors.
The simplest models imply almost no destruction ($<$0.05~dex, possibly 
$\la$0.01~dex) in very metal-poor turnoff
stars$^{\cite{ddk90}}$. Models incorporating
rotationally-induced mixing had predicted large depletion
factors $\sim$~1~dex, though more recent efforts  give lower
values$^{\cite{pwsn98}}$ $\sim$~0.2--0.4~dex, and predict a
range of depletion  factors from star to star. The negligible
intrinsic spread found for very  metal-poor turnoff stars,
$\sigma_{\rm int}~<~0.02$~dex, rules out rotational  depletion
even as low as 0.1~dex$^{\cite{rnb99}}$. As diffusion is also 
absent$^{\cite{rbdt96}}$, we conclude that {\it in situ}
depletion is minor,
$<$0.1~dex, and possibly as little as $\sim0.01$~dex. 

{ Abundance Analyses:}
A Li abundance is derived via a parameter- and model-dependent analysis of a 
stellar spectrum, and systematic uncertainties propagate through to 
$A$(Li)$_p$. Effective-temperature calibrations can differ by up to
150--200~K, higher temperatures resulting in higher Li abundances by 0.065~dex
per 100~K. The scale initially adopted$^{\cite{rnb99}}$ gives temperatures 
cooler than a more recent calibration$^{\cite{aam96}}$ by on average 120~K. We 
now adjust the abundances (Table~1) to the newer calibration$^{\cite{aam96}}$, 
but note that systematic errors of $\pm$120~K may still exist. 
This is one of the largest contributions to the uncertainty in $A$(Li)$_p$. 
Fortunately, errors in the surface gravity, microturbulence, or damping 
parameters are negligible$^{\cite{rnb99}}$. Concerns about 1-D, 
plane-parallel model atmospheres 
have been reduced by simulations of solar-type granulation$^{\cite{u98}}$ which
show that the Li abundance is underestimated in the 1-D approximation by 
$<$~0.10~dex, and possibly $<$~0.01~dex, depending on the theoretical 
prescription for microturbulence. Consistent results in the metal-poor star 
HD~140283$^{\cite{bm98}}$ from the Li~6104~\AA\ and 6707~\AA\ lines inspire 
further confidence. However, models with greater convective flux can lead to Li 
abundances higher by 0.08~dex$^{\cite{rbdt96}}$. Corrections for 
non-LTE$^{\cite{crbs94}}$ are only $-0.01$ to $-0.03$~dex, and the uncertainty 
in the $gf$-values is only 0.02~dex (1$\sigma$)$^{\cite{t94}}$.

{ Anomalous objects:}
Apart from the grossly Li-depleted star G186-26, only one of the remaining 
22 objects in our sample was rejected by outlier-detection algorithms, changing
the mean abundance by only $\sim~0.005$~dex. Similar un-recognised objects would
affect the result by $^<_\sim$~0.01~dex.

We can use the primordial element abundances to fix the one free parameter of 
the standard hot Big Bang nucleosynthesis (BBN) 
model$^{\cite{st98,fkot}}$, the baryon-to-photon ratio, $\eta$. From this, 
the baryon density of universe, $\Omega_b$, may be deduced (for a given 
Hubble parameter, $h$). The inferred primordial abundance for Li (Table 1)
is
$A$(Li)$_p$~= 2.09$^{+0.19}_{-0.13}$
(Li/H~=~1.23$^{+0.68}_{-0.32}\times 10^{-10}$),
where the errors incorporate statistical (negligible) and systematic (more 
significant) effects. These errors are now sufficiently small that the range of
corresponding $\eta$ values (see below) is dominated by the uncertainties in the
input nuclear cross-sections used BBN {\it calculations} rather than in the 
abundance. (Uncertainties in BBN give rise to a range of $\eta$ values 
similar to those quoted below even for
a perfectly determined value of $A$(Li)$_p$.) 

An important test of the BBN model is whether the inferred primordial abundances
give concordant values of $\eta$. This is best tested by establishing likelihood
distributions
(as a function of $\eta$) for each element, convolving the theoretical and 
observational uncertainties$^{\cite{fkot,fo96}}$. Figure~2 shows the 
likelihood distributions for $^4$He and four possible values of the 
primordial \li7 abundance, all of which give excellent agreement. 
Overall concordance (at 95\% CL) occurs for $\eta_{10} \equiv 10^{10}\eta =$ 
(1.4--4.9), (1.5--4.4), (1.7--3.9), and (1.8--3.6), for 
$10^{10}\times ^7$Li/H = 1.9, 1.6, 1.23, and 0.9 respectively.
We can then use this result to assess the diverse deuterium values.
For high D/H (2.0$\times 10^{-4}$), the peak of the D/H likelihood function (not
shown) is at $\eta_{10}$ = 1.7, (95\% CL = 1.4--3.8), in very
good agreement with the results from $^4$He and $^7$Li. For low 
D/H (3.4$\times 10^{-5}$), the peak of the D/H likelihood function is at
$\eta_{10}$ = 5.2 (95\% CL = 4.6--6.1), which
would require $^7$Li$_p$ at the upper end of the possible range. 
However, if the low D/H was even only slightly higher, at 
5$\times 10^{-5}$$^{\cite{ltb}}$, then the D/H peak occurs
at $\eta_{10}$ = 4.0 (95\%\ CL = 3.6--4.6), consistent with the ranges
for $^4$He and $^7$Li. 

The baryon density corresponding to $\eta$ = (1.7--3.9)$\times 10^{-10}$ is 
$\Omega_b = $(0.025--0.057)/$h^2_{50}$ (where $h_{50}$ is the Hubble constant
in units of 50 km s$^{-1}$ Mpc$^{-1}$), significantly below the lower
95\%\ CL on $\Omega_m~=~0.25^{+0.18}_{-0.12}$$^{\cite{eblhe99}}$, thus
maintaining the requirement for non-baryonic dark matter.

{\bf Acknowledgements.}
The authors gratefully acknowledge discussions with 
Drs C. P. Deliyannis, M. Pinsonneault, and J. A. Thorburn
on issues of stellar depletion.
S.G.R. thanks the Institute of Astronomy of the University of Cambridge for
provision of facilities following the closure of the Royal Greenwich 
Observatory.  The work of KAO was supported in part by the Department of
Energy at the University of Minnesota. 

Correspondence 
should be addressed to S.G.R.  
(e-mail: s.g.ryan@open.ac.uk)

\vfill
\eject



\vfill
\eject

\begin{center}
\begin{tabular}{llll}
\hline
\multicolumn{4}{c}{\bf Table 1 Inferred Primordial Lithium Abundance}\\
\multicolumn{1}{l}{-} &&& \multicolumn{1}{r}{-} \\
Observed:$^{\cite{rnb99}}$	
	&\multicolumn{1}{r}{$\langle A$(Li)$\rangle _{-2.8}$ =}
						&+2.12	&$\pm$0.02\\
\multicolumn{1}{l}{-} &&& \multicolumn{1}{r}{-} \\
\multicolumn{4}{l}{Corrections to apply (logarithmic):} (Estimated\\ 
\multicolumn{4}{l}{(1) GCE/GCR:} Uncertainty)\\ 
~~~~~previous analyses$^{\cite{rnb99}}$   &$-0.14$ to $-0.05$\\
~~~~~log data fit (eq. (1))		   &$-0.20$ to $-0.09$\\	
~~~~~linear data fit (eq. (2))		   &$-0.12$ to $-0.04$\\	
~~~~~linear data fit (eq. (3))		   &$-0.16$ to $-0.05$\\	
~~~~~model fits (eq. (2)--(3))	   	   &$-0.12$ to $-0.02$\\
~~~~~Adopted (excludes model):   &	 &$-0.11$&$^{+0.07}_{-0.09}$\\
\multicolumn{2}{l}{(2) Stellar depletion			}&$+0.02$&$^{+0.08}_{-0.02}$\\
\multicolumn{2}{l}{(3a) $T_{\rm eff}$ scale zeropoint		}&$+0.08$&$\pm$0.08\\
\multicolumn{2}{l}{(3b) 1-D atmosphere models			}&$+0.00$&$^{+0.10}_{-0.00}$\\
\multicolumn{2}{l}{(3c) Convective treatment			}&$+0.00$&$^{+0.08}_{-0.00}$\\
\multicolumn{2}{l}{(3d) NLTE					}&$-0.02$&$\pm$0.01\\
\multicolumn{2}{l}{(3e) $gf$-values				}&$+0.00$&$\pm$0.04\\
\multicolumn{2}{l}{(4) Anomalous objects			}&$+0.00$&$\pm$0.01\\
\ \\
Total						&&$-0.03$&$^{+0.19}_{-0.13}$\\
\multicolumn{1}{l}{-} &&& \multicolumn{1}{r}{-} \\
Inferred:	& \multicolumn{1}{r}{$A$(Li)$_p$ =}		 &+2.09   &$^{+0.19}_{-0.13}$\\
\hline
\end{tabular}
\end{center}

{\bf Table Caption:}

Table 1: 
Weighted mean  and 95\% CL uncertainty of observed Li abundances for a
very metal-poor turnoff  sample$^{\cite{rnb99}}$ with
$\langle$[Fe/H]$\rangle$~=~$-2.8$, and corrections required to deduce the
primordial value. (The {\it weighted} mean differs  slightly from the
{\it robust} mean, viz. 2.11$^{\cite{rnb99}}$.) Five estimates of the
logarithmic correction for GCE are listed, based on the previous 
analysis$^{\cite{rnb99}}$ (logarithmic fits and observed $^6$Li/$^7$Li
ratio)  and the new work in this paper (linear fits and various GCE
models). The model fits are based on the model slopes and the inferred
deviation of the primordial value to the weighted mean at [Fe/H] = -2.8.
The various error estimates, which include random and systematic
uncertainties, are clearly non-Gaussian, so combining them is an
imprecise and subjective  process. We take quadratic sums for the
positive and negative uncertainties  separately, and regard these as
estimates of the 95\% confidence limits.

\vfill
\eject

\vfill
\eject

{\bf Figure captions}
 
\label{fig:Li_sources}
{\bf Fig. 1 } 
Contributions to the total predicted lithium abundance from the adopted GCE 
model$^{\cite{fo99,fo99na}}$, 
compared with low metallicity$^{\cite{rnb99}}$ and
high metallicity$^{\cite{lhe91}}$ stars. The solid curve is the sum of all 
components. 
Additional stellar production mechanisms of \li7 are required for stars with 
[Fe/H] near zero, but are expected to be unimportant for the lowest metallicity
objects. The primordial component of \li7 decreases at high metallicity due to 
astration, but other components increase with metallicity as described in the 
text. The GCE model$^{\cite{fo99,fo99na}}$ is a one-zone (closed box) model 
which includes GCR and stellar nucleosynthesis. The latter does not reproduce 
the observed O-Fe scaling$^{\cite{isr,boes}}$, so we retain the calculated O 
evolution and force Fe to match $[{\rm O/Fe}] = \dofe \feh$, (with 
$\dofe = -0.31$). For GCR production of Li, the model assumes that the cosmic 
ray flux is proportional to the supernova rate, and that GCR abundances match 
the ISM. These assumptions fix the linearity of Li-O scaling at low metallicity
where $\alpha+\alpha$ dominates.
The scale {\it factor} is set by the GCR particle spectrum and confinement, for
which we take a source spectrum $\propto p^{-2}$, and an escape path-length 
$\Lambda = 100 \, {\rm g} \, {\rm cm}^{-2}$. The model requires three remaining 
inputs:
(1) an adopted primordial \lip\ abundance,
(2) the overall normalization for all GCR production, and 
(3) the \nupro\ contribution.
As GCR nucleosynthesis also produces beryllium and boron, we use the meteoritic
\be9 and \b10 abundances to establish to overall normalization, which then fixes
the GCR contributions to \li6 and \li7. 
With these normalizations, the modeled evolution of Be and B fits available 
Population II observations$^{\cite{fo99,fo99na}}$.

\label{fig:likelihood}
{\bf Fig. 2 } 
Likelihood distributions for four values of primordial $^7$Li/H 
($10^{10}\times ^7{\rm Li/H} = 1.9$ ({\it dashed}), 1.6 ({\it dotted}), 
1.23 ({\it solid}), and 0.9 ({\it dash-dotted})), 
and for $^4$He ({\it shaded}) for which we adopt $Y_p = 0.238\pm 0.002\pm 0.005$
(random and systematic uncertainties)$^{\cite{fo98}}$. 
For $^7$Li/H = 1.6$\times 10^{-10}$ ($A$(Li)~=~2.20), there are two likely 
values of $\eta_{10} \equiv 10^{10}\eta$ = 1.9 and 3.6, because the 
predictions are not monotonic in $^7$Li. For $^7$Li/H~$\la~1.1\times
10^{-10}$ ($A$(Li)~$\la$~2.04),
the Li abundance is at or below the BBN predicted value, so there is only one 
peak, at $\eta_{10}$ $\simeq$ 2.6; uncertainties in the prediction and 
observation prevent the likelihood function from vanishing.
The peaks of the combined distribution (the product of $L_{^4{\rm He}}(\eta)$ 
and $L_{^7{\rm Li}}(\eta)$; not shown) 
are at roughly the same value of $\eta$ as in the individual $L_{^7Li}(\eta)$ 
distributions.

\clearpage
\begin{figure}[!htb]
\begin{center}
\leavevmode
\epsfxsize=160mm
\epsfbox{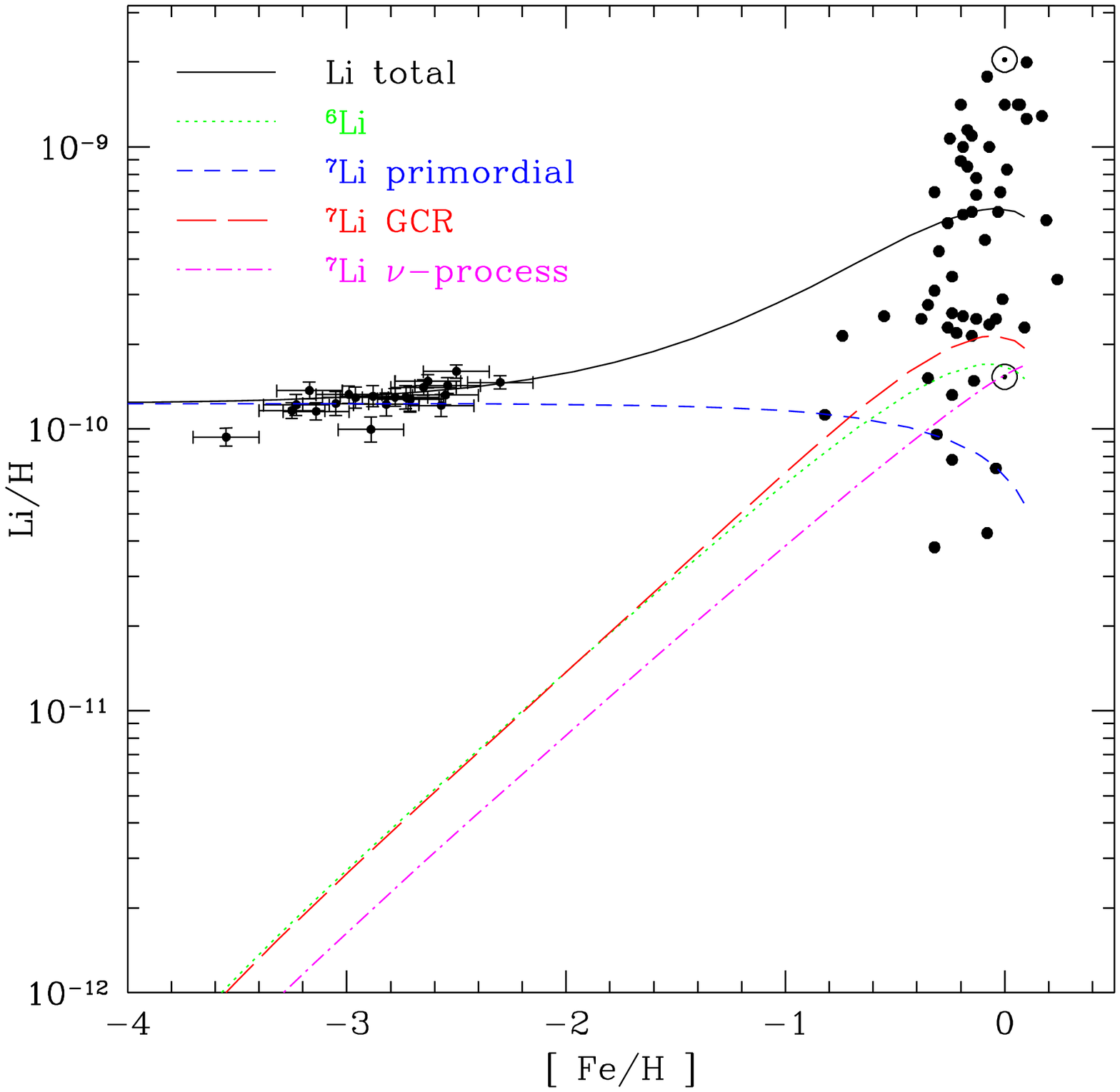}
\vskip .3in
Figure 1
\end{center}
\end{figure}

\clearpage
\begin{figure}[!htb]
\begin{center}
\leavevmode
\epsfxsize=160mm
\epsfbox{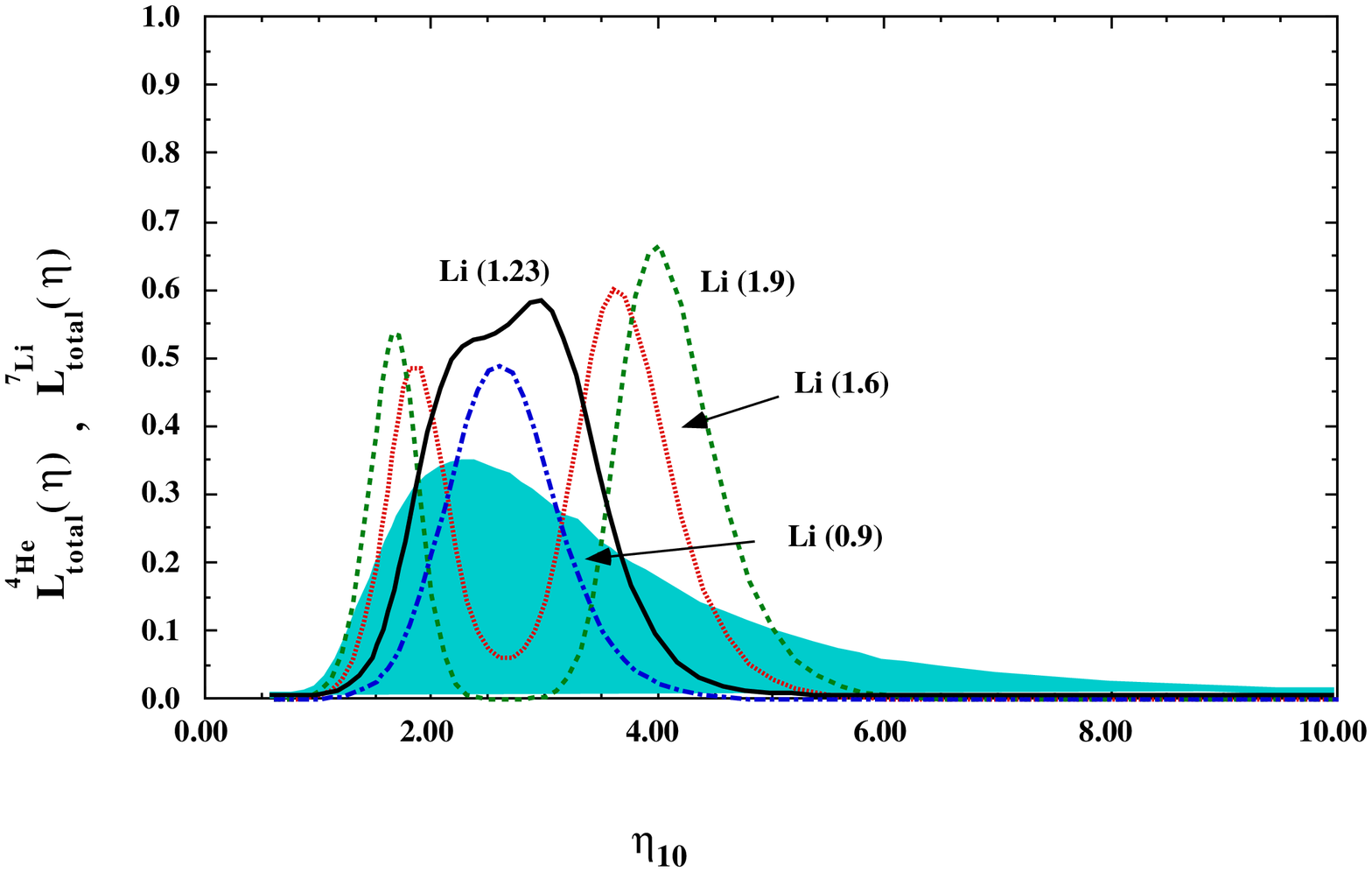}
\vskip 1in
Figure 2
\end{center}
\end{figure}

\end{document}